\documentclass[10pt]{iopart}

\usepackage[T1]{fontenc}

\usepackage{graphicx}
\usepackage{hyperref}
\usepackage{subcaption}
\usepackage{url}
\usepackage[normalem]{ulem}
\usepackage{iopams}
\usepackage{xcolor}
\usepackage{ulem}

\expandafter\let\csname equation*\endcsname\relax
\expandafter\let\csname endequation*\endcsname\relax

\usepackage{amsmath}

\begin{document}

\title{3D simulations of negative streamers in CO$_2$ with admixtures of C$_4$F$_7$N}
\author{Thomas J.G. Smits$^1$, Jannis Teunissen$^{1,3}$, Ute Ebert$^{1,2}$}

\address{
  $^1$ Centrum Wiskunde \& Informatica, Amsterdam, The Netherlands\\
  $^2$ Eindhoven University of Technology, Eindhoven, The Netherlands\\
  $^3$ Centre for mathematical Plasma Astrophysics, KU Leuven, Belgium}

\ead{thomas.smits@cwi.nl}

\begin{abstract}
  CO$_2$ with an admixture of C$_4$F$_7$N could serve as an eco-friendly alternative to the extreme greenhouse gas SF$_6$ in high-voltage insulation.
  Streamer discharges in such gases are different from those in air due to the rapid conductivity decay in the streamer channels.
  Furthermore, since no effective photoionisation mechanism is known, we expect discharge growth to be more stochastic than in air.
  In this paper we investigate whether conventional fluid models can be used to simulate streamers in CO$_2$ with admixtures of C$_4$F$_7$N of 1 or 10\%.
  We focus on 3D simulations of negative streamers.
  First we review cross section databases for C$_4$F$_7$N and CO$_2$.
  Then we compare a two-term Boltzmann solver with a Monte Carlo method to compute reaction and transport coefficients from the cross sections.
  Afterwards we compare 3D fluid simulations with the local field (LFA) or local energy approximation (LEA) against particle simulations.
  In general, we find that the results of particle and fluid models are quite similar.
  One difference we observe is that particle simulations are intrinsically stochastic, leading to more branching.
  Furthermore, the LEA model does not show better agreement with the particle simulations than the LFA model.
  We also discuss the effect and choice of different boundary conditions on the negative rod electrode.

\end{abstract}
\ioptwocol

\section{Introduction}
\label{section:chapter1}
\subsection{Streamers in air and in new insulation gases}
Streamers are ionisation fronts with strong local field enhancement. 
They frequently determine the initial stage of electrical breakdown of gases \cite{nijdam2020physics}.
Streamers in air are the most widely investigated, but their behaviour does not seem to be generic for other gases because of the strong photoionisation mechanism in air.
In this study, we analyse simulations of negative streamers in mixtures of CO$_2$ with C$_4$F$_7$N (abbreviated as CFN), which is an eco-friendly insulation gas \cite{rabie2018assessment}.

$\textrm{SF}_6$ is used as an insulation gas in many high-voltage applications, but it is the most potent greenhouse gas \cite{IPCC2021, UNFCCC_gwp}. 
Hence, SF$_6$ urgently needs to be replaced.
The large electron attachment coefficient and, therefore the high critical field make CFN-CO$_2$ mixtures suitable candidates for high-voltage insulation. 
However, streamer discharges in eco-friendly alternatives differ significantly from those in air, as shown in \cite{skirbekk2022pre} for C5-Fluoroketone with air. 
Similar behaviour has been observed in \cite{guo20233d} and in \cite{vemulapalli2024measurement} for CO$_2$ with admixtures of CFN. 
On the microscopic level, these gases have high attachment rates and an unknown mechanism of photoionisation.
On the macroscopic level, the discharges in the insulation gases do not form long conducting streamer channels. 
Instead, there are isolated streamer heads that only propagate if the background field is close to the critical field \cite{guo20233d}. 
These macro- and microscopic properties differ significantly from discharges in air.

Another distinction lies in the propagation of positive versus negative streamers. Negative streamers have a negatively charged layer of electrons around their head, and they propagate against the electric field. 
Electrons are accelerated away from the streamer, so no external source of free electrons ahead of the streamer is needed. 
In contrast, positive streamers have a positively charged layer at their head and move along the electric field. 
Electrons are accelerated toward them, requiring a source of free electrons ahead of the streamer for propagation \cite{nijdam2020physics}. Therefore, negative streamers can propagate in any gas, but the existence of positive streamers depends on a source of free electrons. For air photoionisation provides such a source, but in most other gases such a mechanism is unknown.

In~\cite{seeger2016streamer} it was observed that negative CO$_2$ streamers were faster than positive streamers in strongly non-uniform background fields. 
We expect similar behaviour in CO$_2$ with an admixture of CFN, when CO$_2$ is the main component. 
It is unclear if efficient free electron sources exist in front of the streamers in CFN-CO$_2$ mixtures.
If this is not the case, negative streamers could propagate more easily than positive ones.
This would be opposite to the situation in air, in which positive streamers are dominant \cite{nijdam2020physics}.

\subsection{Publications on 2D fluid simulations}
\label{sec:12}
Streamer discharges are commonly simulated with a fluid model using the local field approximation (LFA). 
In the LFA it is assumed that electrons instantaneously relax to the local electric field, so that transport and reaction coefficients are functions of the local electric field.
Studies of axisymmetric streamers in CO$_2$ with admixtures of CFN using fluid models and the LFA have been presented in \cite{yan2023surface,vu2020numerical,yan2023numerical,fan2022simulation,gao2022negative,wang2021numerical}.

If electrons pass through a rapidly changing electric field, the validity of the LFA is questionable.
An improvement of this classical approximation is the local energy approximation (LEA), in which transport and reaction coefficients are functions of the local mean electron energy.
Examples in which an LEA is required are given in \cite{grubert2009local,diasAreLocalfieldLocalenergy2025}. 
While the LFA includes one reaction-drift-diffusion equation for the electron density, the LEA requires an additional reaction-drift-diffusion equation for the mean electron energy. 
LFA and LEA can be derived from the velocity moments of the Boltzmann equation with a truncation of the expansion after the first or second term \cite{dujko2013high, markosyan2015comparing}. 
Fewer studies have used the LEA; a study of N$_2$ with an admixture of CFN has been published in~\cite{levko2024computational}.

Many authors have used 2D axisymmetric fluid models to simulate discharges in gases with admixtures of CFN~\cite{yan2023surface,vu2020numerical,yan2023numerical,fan2022simulation,gao2022negative,wang2021numerical,levko2024computational}. 
These models have been used to study the influence of different electrode shapes \cite{yan2023numerical} or the characteristics of partial discharges in gases with an admixture of CFN instead of SF$_6$ \cite{vu2020numerical,yan2023surface,fan2022simulation}.
In \cite{wang2021numerical}, a 2D fluid model was used to investigate the effect of buffer gases (either $\mathrm{N}_2$ or $\mathrm{CO}_2$) with admixtures of CFN, and the results were compared with results of similar $\mathrm{SF}_6$ admixtures.
In \cite{levko2024computational} the effect of the CFN fraction and cathode voltage rise rate on the breakdown voltage was modelled with an axisymmetric model and LEA.

\subsection{Publications on 3D particle simulations}
2D fluid models, both with LFA or LEA, are computationally efficient, but they have important constraints:
\begin{enumerate}
    \item They are axisymmetric by construction.
    The models cannot simulate streamer branching or interaction of several streamers (except if they are aligned along one axis).
    \item They are deterministic without the stochastic fluctuations that are physically present. 
    \item In low-density regions a tiny fraction of a single electron can be present in a numerical grid cell, leading to numerical artifacts.
\end{enumerate}
3D particle simulations as performed in \cite{guo20233d} can overcome these shortcomings.
In this work negative streamers were simulated in 3D with a particle in cell (PIC) Monte Carlo collision method.
Different background fields and different mixtures of CFN and CO$_2$ were considered.
The paper captured the stochastic behaviour of inception and branching, and found that fields close to breakdown are required for streamer propagation.
It was also shown that the negative streamers will evolve as frequently branching streamer heads, with only a short conducting streamer channel.

\subsection{The content of our paper: cross sections, Boltzmann solvers and 3D simulations}

A downside of particle models is their high computational cost.
In the present paper, we therefore investigate whether 3D fluid models can also be used to accurately simulate negative streamers in CO$_2$ with admixtures of CFN.

Another goal is to determine which cross section databases for electron-molecule interactions and which Boltzmann solver type should be used for the input data. 
PIC directly uses cross sections as input, while the fluid models require transport and reaction coefficients. 
These coefficients are calculated from the cross sections using Boltzmann solvers or a Monte-Carlo particle swarm code. 
We study the sensitivity of the coefficients on solver types at different mixture rates of CO$_2$ with CFN.
We also test the different cross section datasets for CO$_2$ and CFN and determine the sensitivity of the reduced transport and reaction coefficients on these cross sections.
The aim is to find a suitable solver and cross section dataset for the 3D fluid model. Then we test different LFA, LEA and PIC models in simulations of negative streamers. 

\subsection{Organisation of the paper}
This work is structured as follows: in section \ref{subsection:chapter3_1}, the transport and reaction coefficients are determined with the different available cross section databases. 
The sensitivity of these profiles to the input cross sections and to the used solver is probed, resulting in a recommendation for the cross section dataset and the Boltzmann solver to use. 
Next, in Chapter \ref{section:chapter2}, the fluid models and the particle model are explained in Sections \ref{subsection:fluidmodel} and \ref{subsection:picmodel}. 
The numerical setup and the initial and boundary conditions are discussed in sections \ref{subsubsection:fluid_numerical_setup}, \ref{subsubsection:fluid_initial_and_boundary_conditions}, \ref{subsection:ic_sensitivity} and \ref{subsection:BC_versus}. 
The results of the fluid and the particle models are compared in sections \ref{subsection:different_bg_fld}, \ref{subsection:compare_diff_voltage}, \ref{subsection:different_concentration} and \ref{subsection:big_domain}. 
The discussion of this work is presented in section \ref{section:chapter4}, and the conclusion is given in section \ref{section:chapter5}.

%%% Local Variables:
%%% mode: LaTeX
%%% TeX-master: "main"
%%% End:

\section{Input data}
\label{subsection:chapter3_1}
\subsection{Electron-neutral cross sections}
\label{sec:electr-neutr-cross}
We test cross section data sets derived by different authors, as collected on the LXCAT database \cite{pancheshnyi2012lxcat, carbone2021data,pitchford2017lxcat}.
For CFN, the databases of Flynn \cite{flynn2023generation} and of Zhang \cite{zhang2023determination} are tested. For CO$_2$, the databases of Hayashi \cite{Hayashi_co2}, Triniti \cite{Triniti_co2}, Phelps \cite{Phelps_co2}, Morgan \cite{Morgan_co2}, and IST-Lisbon \cite{IST_co2} are tested.
The Phelps, Morgan, and IST-Lisbon cross sections for CO$_2$ include an `effective' momentum scattering cross section, which we convert to an elastic cross section by subtracting the sum of all inelastic processes.

\subsection{Comparing transport and reaction solvers}
\begin{figure}
    \centering
    \includegraphics[width=1\linewidth]{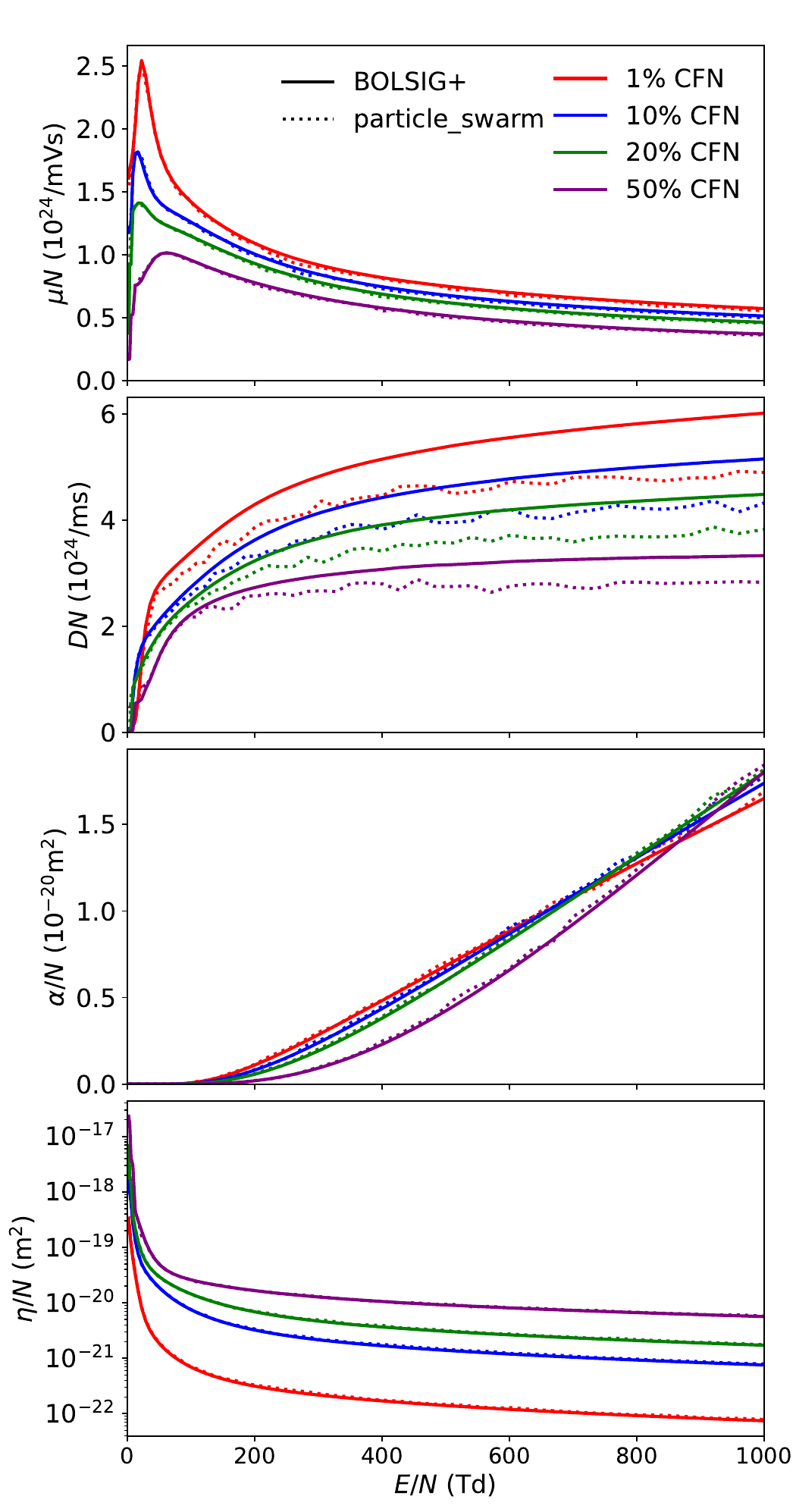}
    \caption{Comparison of the reduced flux transport and reaction coefficients calculated with BOLSIG+ (solid lines) and \texttt{particle\_swarm} code (dotted lines). 
    The coefficients are flux coefficients and were calculated for CO$_2$ with admixtures of $1\%,10\%,20\%$ and $50\%$ CFN as a function of $E/N$. 
    Cross section data of Hayashi for CO$_2$ and Flynn for CFN were used.
    $\mu N$, $\alpha/N$, and $\eta/N$ agree well, while $DN$ differs as discussed in the main text.\\}
    \label{fig:transport_solver}
\end{figure}

Fluid models require electron transport and reaction data, which can be computed from electron-neutral cross sections using a Boltzmann solver or a particle model.
Since electrons attach very quickly to CFN in low fields, we were not sure whether the choice of solver type would affect the resulting transport data.
We therefore tested two solvers: BOLSIG+ \cite{hagelaar2005solving} and the \texttt{particle\_swarm} code \cite{JannisTeunissen_2025}. 
These solvers are used to calculate the reduced transport and reaction data for CO$_2$ with different admixture ratios of CFN.
Reduced coefficients are studied as these are used in the fluid simulations as input.
We used CO$_2$ with admixture of 1, 10, 20 or 50\% CFN at 300 K and 1 bar such that the gas number density is $N=2.41\times10^{25}$ m$^{-3}$.
Figure \ref{fig:transport_solver} shows the calculated reduced transport and reaction data, using the Flynn cross sections for CFN and the Hayashi cross sections for $\textrm{CO}_2$.
We show the flux mobility and the flux transverse diffusion coefficient used in the fluid simulations.

From figure \ref{fig:transport_solver}, we conclude that the two solvers produce similar coefficients for the reduced mobility $\mu N$, ionisation rate $\alpha/N$, and attachment rate $\eta/N$.
To quantify this, for each mixture and transport coefficient we have determined the average relative difference between the solvers.
This was done by first computing relative differences and then taking the average of these relative differences. 
For $\alpha/N$, the average was taken only for fields above 150 Td, to avoid division by zero.
Average relative differences are below $3\%$ for $\mu N$ and $\alpha/N$, and for $\eta/N$ they are below $5\%$.
The most significant deviations occur in low fields with a large fraction of CFN.

Significant deviations are observed for the transverse flux diffusion coefficient shown in figure~\ref{fig:transport_solver}.
Transverse and bulk diffusion coefficients, which are not shown in the figure, also differ significantly between the solvers.
The inaccuracy of the two-term approximation causes these discrepancies between solvers, see e.g.~\cite{e505e7dc8948418a877ec112866d9175, petrovic2009measurement,Stephens_2018a,hagelaar2025beyond}.
We have verified that a multiterm Boltzmann solver with six terms~\cite{Stephens_2018a} gives excellent agreement with the diffusion coefficients obtained with the \texttt{particle\_swarm} code. 
Therefore we trust these two codes and use the \texttt{particle\_swarm} code to calculate the input coefficients for the fluid simulations.

\subsection{Dependence on cross sections}
\label{subsection:comparing_c4}
\begin{figure}
    \centering
    \includegraphics[width=1\linewidth]{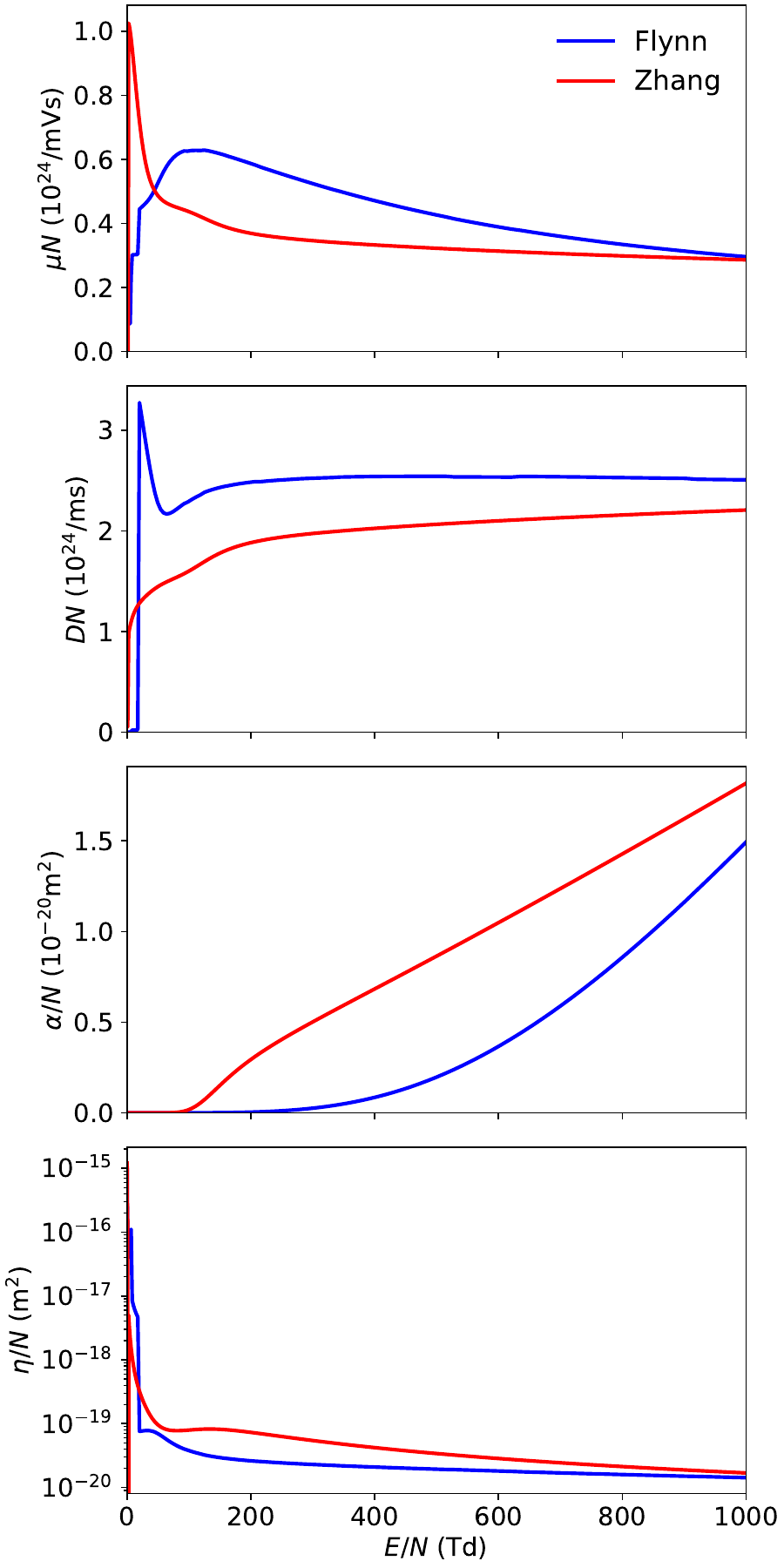}
    \caption{Comparison of the Flynn and Zhang reduced transport and reaction coefficients in pure CFN as functions of $E/N$.
    The coefficients are flux coefficients and were calculated with BOLSIG+.
    Differences of up to $40\%$ for all coefficients are observed.}
    \label{fig:transport_c4}
\end{figure}

We now determine how the choice of cross sections affects reduced transport and reaction data in pure CFN, using BOLSIG+ at $N=2.41\times10^{25}$ m$^{-3}$.
Figure~\ref{fig:transport_c4} shows results in pure CFN, using either the Zhang or the Flynn cross sections.
There are clearly significant differences in $\alpha/N$, $\eta/N$, $\mu N$ and $DN$, for which we have no explanation.
Experimental swarm data for pure CFN is only available between 700 Td and 1050 Td (not shown here).
Hence, it is difficult to conclude which CFN cross section set is to be preferred.

\begin{figure*}
    \centering
    \includegraphics[width=0.9\textwidth]{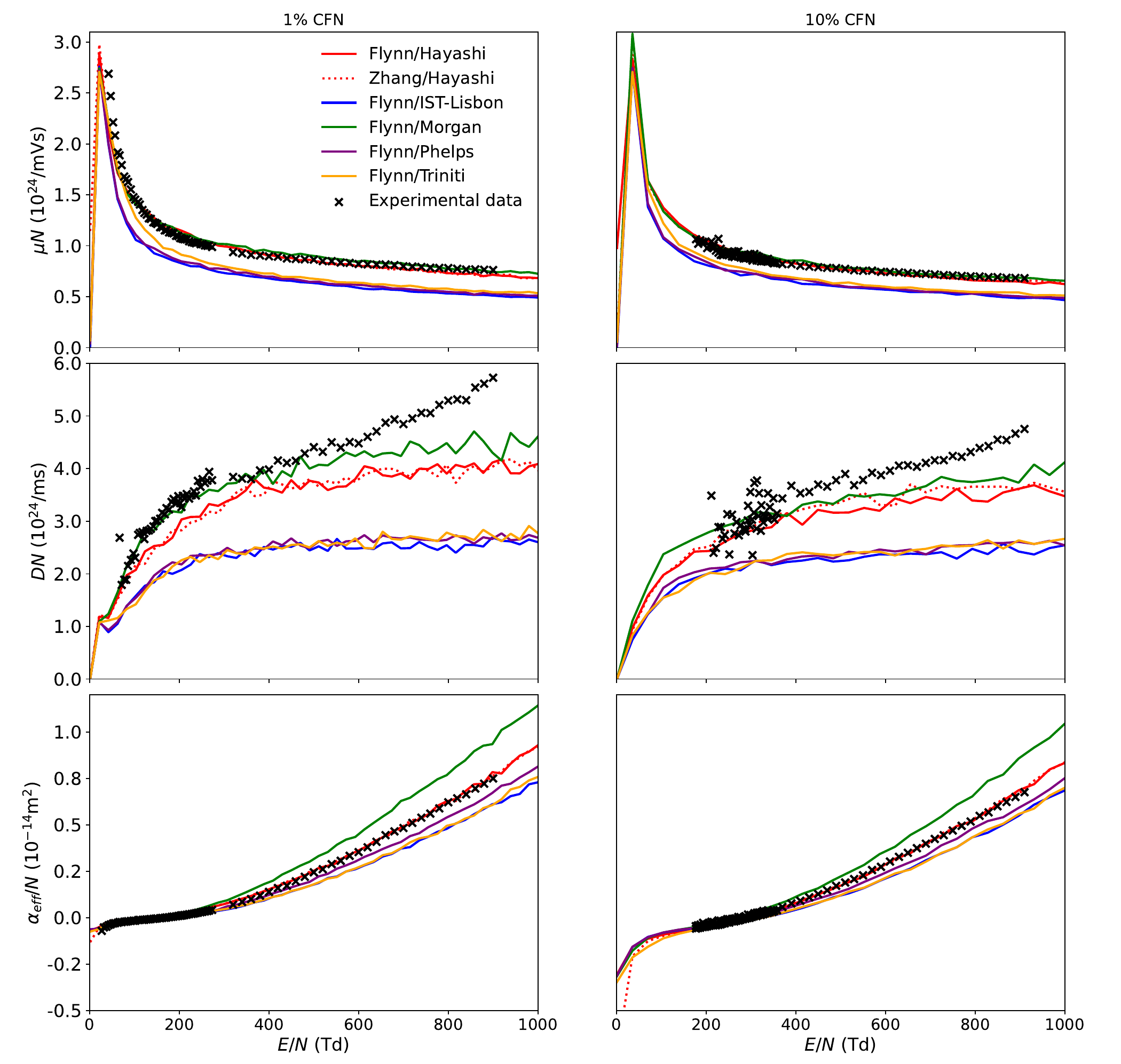}
    \caption{Comparison of reduced transport and reaction coefficients of different CO$_2$ databases with an admixture of 1\% (left) or 10\% (right) CFN as functions of $E/N$.
    The coefficients are bulk coefficients, and were calculated with the \texttt{particle\_swarm} code.
    The black crosses indicate experimental swarm data from \cite{chachereau2018electrical,vemulapalli2023pulsed} at 296 K and 100 Pa. 
    The CFN cross sections here have a minor influence, so for the Zhang database only the combination with the Hayashi CO$_2$ cross section is shown.
    The Flynn cross sections are used to calculate the input data for the fluid simulations. 
    Experimental data agree well with the Flynn/Hayashi and Zhang/Hayashi results.}
    \label{fig:comb_experiment}
\end{figure*}

However, in CO$_2$ with admixtures of CFN, swarm data have been measured over a wide range of reduced electric fields~\cite{chachereau2018electrical,vemulapalli2023pulsed} and these data are available on LXCat~\cite{pitchford2017lxcat}.
In figure~\ref{fig:comb_experiment}, reduced coefficients calculated from different CFN and CO$_2$ cross section datasets are compared with experimental swarm data, for $1\%$ and $10\%$ CFN at $100\;\textrm{Pa}$ and $296\;\textrm{K}$.
For this comparison, we computed the bulk mobility, the bulk longitudinal diffusion coefficient and $\alpha_{\textrm{eff}}/N$ using the \texttt{particle\_swarm} code.
Here $\alpha_{\textrm{eff}} = \alpha - \eta$.
Swarm experiments provide bulk coefficients, so we also calculate bulk coefficients in this section.
Note, flux coefficients are used as input for the simulations.

In contrast to the case of pure CFN, the CFN cross sections of either Zhang or Flynn now result in similar transport data for the considered CFN admixtures in CO$_2$. 
The only exception is that $\alpha_{\textrm{eff}}/N$ becomes significantly more negative with the Zhang cross sections for small $E/N$ with 10\% CFN. 
The coefficients in pure CO$_2$ are given in the appendix for comparison.
Due to the small effect of CFN cross sections, only one set of coefficients based on the Zhang cross section is shown in figure~\ref{fig:comb_experiment}.
Since the mixtures predominantly contain CO$_2$, the transport data are more sensitive to the choice of CO$_2$ cross sections.
For $\mu_{\textrm{bulk}}N$, there is good agreement between the experimental data (black crosses) and data computed using the Triniti and Hayashi databases.
For $\alpha_{\textrm{eff}}/N$, good agreement is only obtained with the Hayashi database.
The agreement between the calculated and the experimental diffusion data is generally worse, in particular for fields above $600\;\textrm{Td}$.
The best agreement between calculated and experimental diffusion data is obtained with the Morgan database, closely followed by the Hayashi database.

Since the electron mobility and the effective ionisation coefficient are the most important input data in a fluid model, we will use the Hayashi CO$_2$ cross sections in our simulations.
The panels in Figure~\ref{fig:comb_experiment} do not show significant differences when using either the Zhang or the Flynn cross sections for CFN, so we have made the rather arbitrary choice to use the Flynn cross sections.
The critical fields $E_k$ for CO$_2$ with CFN admixtures using Hayashi and Flynn cross sections are given in table~\ref{tab:Critical_field}.

\begin{table}
    \centering
    \caption{Critical electric field $E_k$ for CO$_2$ with different admixtures of CFN at $N=2.41\times10^{25}$ m$^{-3}$.
    Hayashi cross sections for CO$_2$ and Flynn cross sections for CFN are used.
    The critical field $E_k$ is defined as the field where the ionisation rate $\alpha$ equals the attachment rate $\eta$.
    }
    \label{tab:Critical_field}
    \begin{tabular}{c c}
    \hline
       \textbf{Admixture of CFN}  &  $\mathbf{E_k}$ \\
    \hline
        pure \;$\mathrm{CO}_2$ & $18.0\;\mathrm{kV/cm}$ \\
        $1\%\;\mathrm{CFN}$ & $35.4\;\mathrm{kV/cm}$ \\
        $10\%\;\mathrm{CFN}$ & $69.3\;\mathrm{kV/cm}$ \\
        $20\%\;\mathrm{CFN}$ & $96.5\;\mathrm{kV/cm}$ \\
        $50\%\;\mathrm{CFN}$ & $157.3\;\mathrm{kV/cm}$ \\
        $100\%\;\mathrm{CFN}$ & $234.7\;\mathrm{kV/cm}$ \\
        \hline
    \end{tabular}
\end{table}

The processes included in these databases are listed in table \ref{table:C4F7N}. To reduce the computational cost,
only the species e$^-$, CO$_2^+$, O$^-$, C$_4$F$_7$N$^-$, C$_4$F$_7$N and CO$_2$ from table \ref{table:C4F7N} are tracked. All other species are collected into either M$^-$, M$^+$ or M based on their charge. 
This does not alter the model's predictions, as no detachment or recombination reactions are considered.

\begin{table}
\centering
\caption{The reactions included in the simulations, excluding excitation reactions.
CFN reactions are from Flynn~\cite{flynn2023generation}, and CO$_2$ reactions are from Hayashi~\cite{Hayashi_co2}.}
\label{table:C4F7N}
\begin{tabular}{l}
\hline
   \textbf{Ionization} \\ \hline
   $e + \textrm{CO}_2 \to e + e + \textrm{CO}_2^{+}$                                                                       \\
   $e + \textrm{C}_4\textrm{F}_7\textrm{N} \to e + e + \textrm{C}_3\textrm{F}_4\textrm{N} + \textrm{CF}_3^{+}$             \\
   $e + \textrm{C}_4\textrm{F}_7\textrm{N} \to e + e + \textrm{CF}_4 + \textrm{C}_3\textrm{F}_3\textrm{N}^{+}$             \\
   $e + \textrm{C}_4\textrm{F}_7\textrm{N} \to e + e + \textrm{F} + \textrm{C}_4\textrm{F}_6\textrm{N}^{+}$                \\
   $e + \textrm{C}_4\textrm{F}_7\textrm{N} \to e + e + \textrm{C}_2\textrm{F}_5 + \textrm{C}_2\textrm{F}_2\textrm{N}^{+}$  \\
   $e + \textrm{C}_4\textrm{F}_7\textrm{N} \to e + e + \textrm{C}_3\textrm{F}_4\textrm{N} + \textrm{F}_2 + \textrm{CF}^{+}$ \\
   $e + \textrm{C}_4\textrm{F}_7\textrm{N} \to e + e + \textrm{C}_2\textrm{F}_3\textrm{N} + \textrm{C}_2\textrm{F}_4^{+}$      \\
   $e + \textrm{C}_4\textrm{F}_7\textrm{N} \to e + e + \textrm{C}_3\textrm{F}_5\textrm{N} + \textrm{CF}_2^{+}$             \\
   $e + \textrm{C}_4\textrm{F}_7\textrm{N} \to e + e + \textrm{C}_2\textrm{F}_6 + \textrm{C}_2\textrm{FN}^{+}$             \\ \hline

   \textbf{Attachment} \\ \hline
   $e + \textrm{CO}_2 \to \textrm{CO} + \textrm{O}^{-}$                                                \\
   $e + \textrm{C}_4\textrm{F}_7\textrm{N} \to \textrm{C}_4\textrm{F}_7\textrm{N}^{-}$                 \\
   $e + \textrm{C}_4\textrm{F}_7\textrm{N} \to \textrm{F} + \textrm{C}_4\textrm{F}_6\textrm{N}^{-}$    \\
   $e + \textrm{C}_4\textrm{F}_7\textrm{N} \to \textrm{CF}_3 + \textrm{C}_3\textrm{F}_4\textrm{N}^{-}$ \\
   $e + \textrm{C}_4\textrm{F}_7\textrm{N} \to \textrm{C}_4\textrm{F}_6\textrm{N} + \textrm{F}^{-}$    \\
   $e + \textrm{C}_4\textrm{F}_7\textrm{N} \to \textrm{C}_3\textrm{F}_7 + \textrm{CN}^{-}$             \\
   $e + \textrm{C}_4\textrm{F}_7\textrm{N} \to \textrm{C}_4\textrm{F}_6\textrm{N} + \textrm{F}^{-}$    \\
   $e + \textrm{C}_4\textrm{F}_7\textrm{N} \to \textrm{F} + \textrm{C}_4\textrm{F}_6\textrm{N}^{-}$    \\ \hline

   \textbf{Elastic} \\ \hline
   $e + \textrm{CO}_2 \to e + \textrm{CO}_2$                                           \\
   $e + \textrm{C}_4\textrm{F}_7\textrm{N} \to e + \textrm{C}_4\textrm{F}_7\textrm{N}$ \\
\hline
\end{tabular}
\end{table}

%%% Local Variables:
%%% mode: LaTeX
%%% TeX-master: "main"
%%% End:

\section{Simulation models}
\label{section:chapter2}
\subsection{Fluid models: LFA and LEA}
\label{subsection:fluidmodel}
In this work, two types of reaction-drift-diffusion fluid models for electrons are considered~\cite{markosyan2015comparing, dujko2013high}: the local field approximation (LFA) and the local energy approximation (LEA).
The LFA model is commonly used, but it may fail to capture non-local effects~\cite{li2007deviations,petrovic2007kinetic}.
The LFA model used here is given by
\begin{align} \label{eq:LFA}
  &\partial_t n_e + \nabla \cdot \mathbf{\Gamma}_e = S_e, \\
  &\partial_t n_{\textrm{ion}} = S_e, \\
  &\mathbf{\Gamma}_e = -\mu_e(E)\textbf{E}n_e - D_e(E)\nabla n_e,\\
  &S_e = n_e \mu_e(E) E \alpha_\mathrm{eff}(E).
\end{align}
Here, $n_e$ is the electron density, $n_{\textrm{ion}}$ is the density of positive minus negative ions, $\textbf{E}$ is the electric field and $E$ its magnitude, $S_e$ is the source term due to ionisation and attachment, and $\mathbf{\Gamma}_e$ is the electron flux.
Note that ions are assumed to be immobile.

The LEA model used here is given by:
\begin{align} \label{eq:LEA}
     &\partial_t n_e + \nabla \cdot \mathbf{\Gamma}_e =  S_e, \\
     &\partial_t n_{\textrm{ion}} = S_e, \\
     &\partial_t (n_e \epsilon) + \nabla \cdot \mathbf{\Gamma}_\epsilon = S_\epsilon, \\
     &\mathbf{\Gamma}_e = -\mu_e(\epsilon)\textbf{E}n_e - D_e(\epsilon)\nabla n_e,\\
     &\mathbf{\Gamma}_\epsilon = -\frac{5}{3} \left[\mu_e(\epsilon)\textbf{E} n_e\epsilon + D_e(\epsilon)\nabla (n_e\epsilon)\right],\\
     &S_e = n_e \mu_e(\epsilon) E \alpha_\mathrm{eff}(\epsilon),\\
     &S_\epsilon = -\textbf{E} \cdot \mathbf{\Gamma}_e - P_\mathrm{loss}.
\end{align}
Here $\epsilon$ is the local average electron energy, $(n_e \epsilon)$ is the electron energy density, $\mathbf{\Gamma}_\epsilon$ is the energy flux, $S_\epsilon$ is the energy source term.
$P_\mathrm{loss}$ is the electron energy loss rate due to (in)elastic collisions
\begin{equation}
  \label{eq:energy-loss}
  P_\mathrm{loss} = \mu_e(\epsilon) E(\epsilon)^2,
\end{equation}
where $E(\epsilon)$ is the electric field strength corresponding to the mean energy $\epsilon$ in the input data.
This expression is based on the observation that for an electron swarm the average energy gain should equal the average energy loss, see e.g.\ the documentation of BOLSIG+~\cite{hagelaar2005solving}.

The fluid models are coupled with the electric field, computed in the electrostatic approximation as
\begin{align}
  \nabla^2\phi &= -\frac{\rho}{\epsilon_0} \\
  \textbf{E} &= - \nabla \phi,
\end{align}
where $\rho = e(n_{\rm ion}-n_e)$ is the charge density, $\epsilon_0$ is the vacuum permittivity, and $\phi$ is the electric potential.
For both models, the afivo-streamer code \cite{teunissen2018afivo, Teunissen_2017} is used.
This paper presents the first use of the LEA model within the afivo-streamer code.

With the LFA model, unphysical effects due to electron diffusion against the electron drift direction can occur~\cite{Soloviev_2009,teunissen2020improvements}.
This can cause electrons to diffuse into the sheath near a negative electrode, where the field and therefore $\alpha_\mathrm{eff}$ are high, resulting in a rapid growth of the electron density.
Therefore, a correction factor for the LFA electron source term is used
\begin{equation} \label{eq:source_correction}
  f_e = 1 - \frac{\hat{\textbf{E}} \cdot \mathbf{\Gamma}_{\textrm{diffusion}}}{\|\mathbf{\Gamma}_{\textrm{drift}}\|},
\end{equation}
where $\mathbf{\Gamma}_{\textrm{drift}}$ and $\mathbf{\Gamma}_{\textrm{diffusion}}$ are the drift and diffusive components of the electron flux and $\hat{\mathbf{E}}$ is the electric field unit vector, see~\cite{Soloviev_2009,teunissen2020improvements} for details.

\subsection{Particle model}
\label{subsection:picmodel}
We use a particle-in-cell (PIC) Monte Carlo collision model as implemented in the afivo-pic code \cite{teunissen20163d},
and also used in \texttt{particle\_swarm}.
Electrons are tracked as particles, and ions as immobile densities.
The electron velocity and position are updated with the velocity Verlet scheme.
The gas is modelled as a background that electrons stochastically collide with using the null-collision method.
The input for the PIC model are the electron-neutral cross sections. 
We assume that electron scattering after collisions is isotropic. 
For computational efficiency, electrons are grouped into super-particles with adaptive weights.
The weights are changed so that the number of particles per cell roughly stays below 100,
see~\cite{teunissen20163d} for more details.

\subsection{Computational domain and grid refinement}
\label{subsubsection:fluid_numerical_setup}
\begin{figure}
        \centering
        \includegraphics[width=0.85\linewidth]{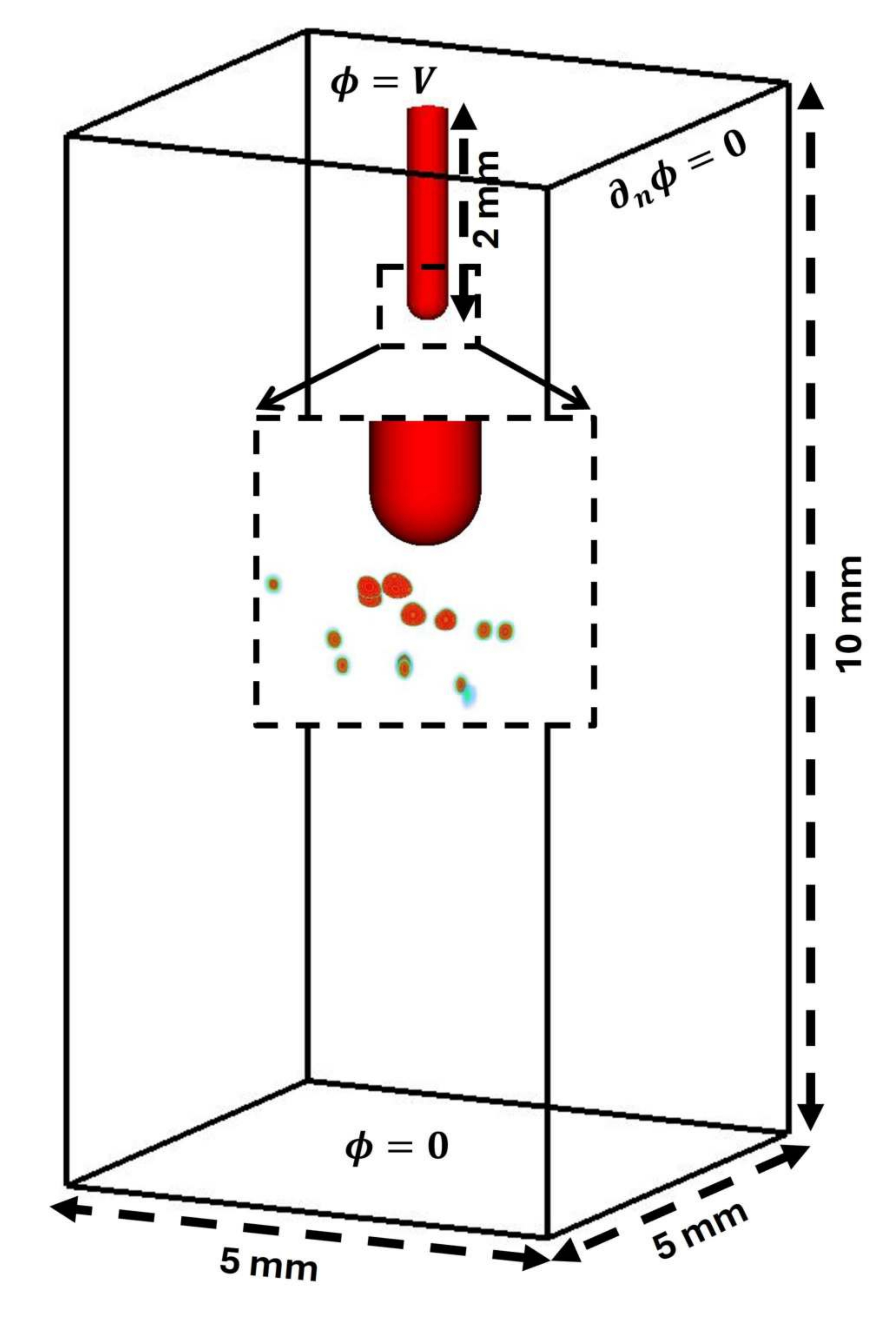}
        \caption{A view of the plate-to-plate computational domain of $5\;\mathrm{mm}\times5\;\mathrm{mm}\times10\;\mathrm{mm}$ with protruding top electrode of $2\;\mathrm{mm}$. The dashed box zooms into the location of the stochastic initial condition after the first time step of $0.2\;\mathrm{ns}$.}
        \label{fig:setup}
\end{figure}

We use a $5\;\textrm{mm}\times5\;\textrm{mm}\times10\;\textrm{mm}$ computational domain with a plate-to-plate geometry and a protruding rod electrode at the top, see figure \ref{fig:setup}; the same configuration was also used in~\cite{guo20233d}.
The rod electrode is $2\;\textrm{mm}$ long and has a semi-spherical tip with a diameter of $0.4\;\textrm{mm}$.
This relatively narrow and short domain reduces the computational cost of the simulations, but we also provide simulations in a larger computational domain in section~\ref{subsection:big_domain}. 

We evaluate the models at $N=2.41\times10^{25}$ m$^{-3}$ with varying admixtures of CFN. 
The simulations use the adaptive mesh refinement of the Afivo model \cite{teunissen2018afivo}. 
The refinement condition is $\alpha_{\textrm{eff}}\Delta x \leq 1.0$ with $\alpha_{\textrm{eff}}$ the effective ionization constant.
This results in a minimal grid spacing of about $2.4\;\mu\textrm{m}$ for the simulations presented here.
The maximum grid spacing is limited to $\Delta x \leq 55\;\mu\textrm{m}$.

\subsection{Initial and boundary conditions}
\label{subsubsection:fluid_initial_and_boundary_conditions}

At the planar bottom boundary, we set $\phi=0$, and at the top boundary, including the protruding electrode, $\phi = V_0$, where $V_0$ is the applied (negative) voltage.
We use the same definition of the background field as in \cite{guo20233d}, so that $E_{bg} = |V_0|/d_{\textrm{plate}}$ with $d_{\textrm{plate}} = 10\;\textrm{mm}$.
Neumann zero boundary conditions ($\textbf{n}\cdot\nabla\phi=0$, where $\textbf{n}$ is perpendicular to the boundary) are applied on the sides of the domain.
For the ion and electron densities, a Dirichlet zero boundary condition ($n_e = 0$) is used on all domain boundaries including the electrodes.
The effect of Neumann and Dirichlet conditions on the species is discussed in section \ref{subsection:BC_versus}.

We sample 15 positions from a Gaussian distribution centred $0.2\;\textrm{mm}$ below the electrode tip with a width of $0.2\;\textrm{mm}$, and we place electron-$\textrm{CO}_2^+$ pairs at these locations as an initial condition. 
In the fluid models the electron-ion pairs are converted to densities.
For better comparison between the models the same 15 initial positions are used in all simulations performed with the particle and fluid models; they are shown in figure~\ref{fig:setup}.

%%% Local Variables:
%%% mode: LaTeX
%%% TeX-master: "main"
%%% End:

\section{Fluid versus particle models}
\label{section:chapter3_2}
\subsection{Comparison of the three models at 36.1 kV/cm}
\label{subsection:different_bg_fld}
\begin{figure}
    \centering
    \includegraphics[width=1\linewidth]{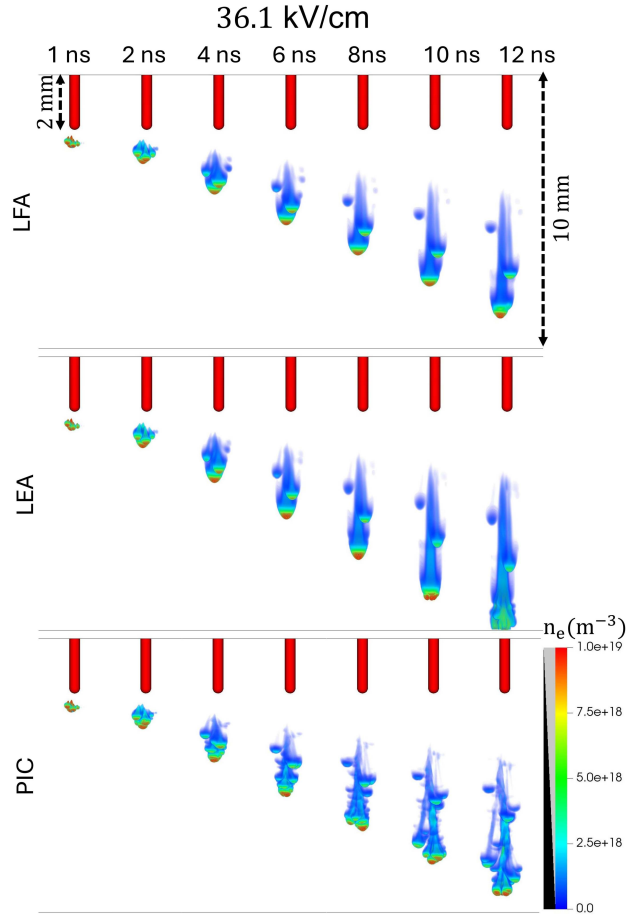}
    \caption{Evolution of the electron density of the LFA, LEA and PIC models for CO$_2$ with an admixture of $1\%$ CFN in a background field of $E_{\mathrm{bg}} = 36.1\;\mathrm{kV/cm}$.
    The simulations show the same initial stochastic growth, with some disagreement in later phases due to branching.
    Figures are generated using ray-traced 3D volume rendering in Visit \cite{VISIT}.}
    \label{fig:36kV}
\end{figure}

\begin{figure}
    \centering
    \includegraphics[width=1\linewidth]{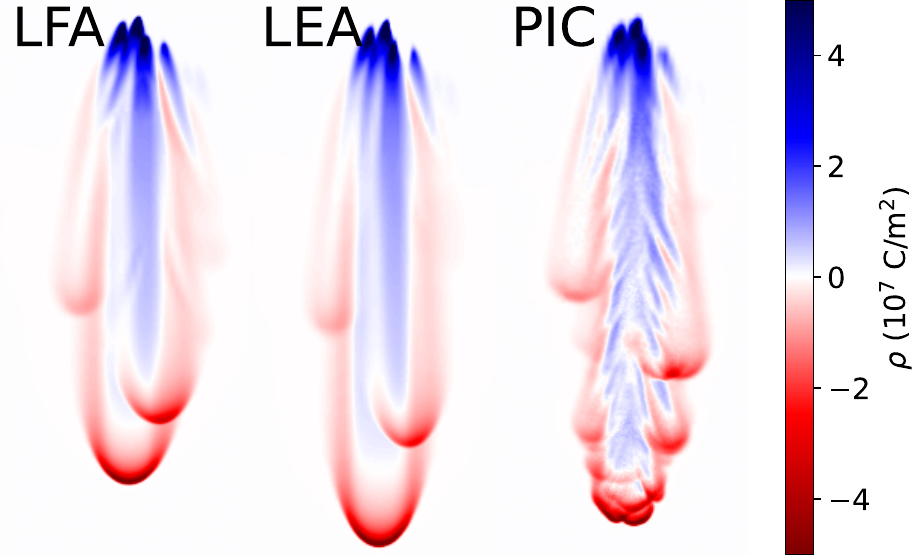}
    \caption{Projection of the charge density $\rho$ on the $xz$-plane at time $8\;\mathrm{ns}$, showing again the simulation data of figure~\ref{fig:36kV}. 
    Red indicates negative and blue positive charge densities.
    The profiles are similar, with the main difference that there is more branching in the particle model.
    }
    \label{fig:36kV_rho_all}
\end{figure}

\begin{figure}
    \centering
    \includegraphics[width=1\linewidth]{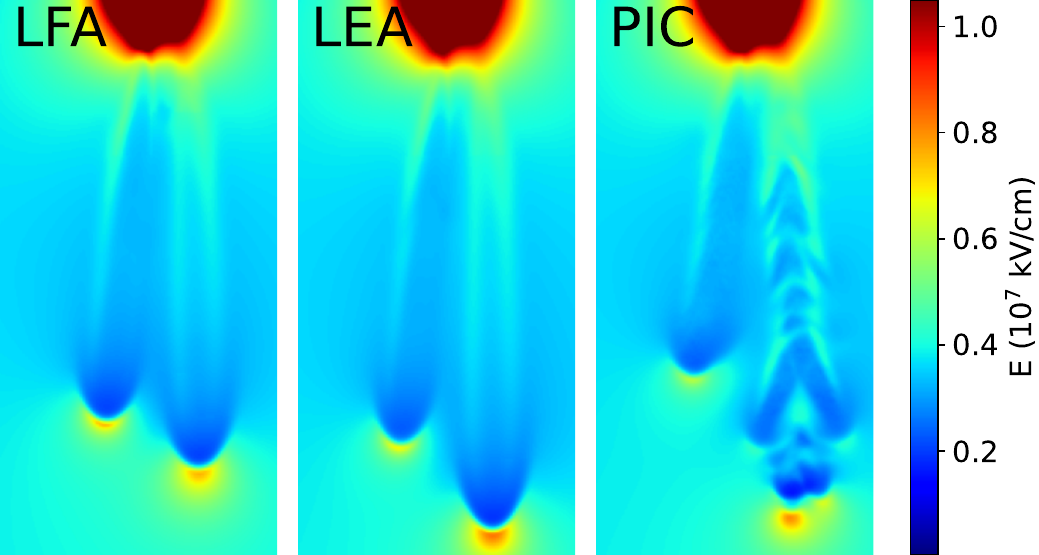}
    \caption{The electric field distribution in the $yz$-plane at time $8\;\mathrm{ns}$, showing again the simulation data from figure~\ref{fig:36kV}. 
    The saturated red region indicates the high field near the electrode. 
    The electric field distributions are similar in the streamer heads for the three models.
    The major difference is again the branching.}
    \label{fig:36kV_electric_fld}
\end{figure}

\begin{figure*}
    \centering
    \begin{subfigure}{0.48\linewidth}
        \centering
        \includegraphics[width=\linewidth]{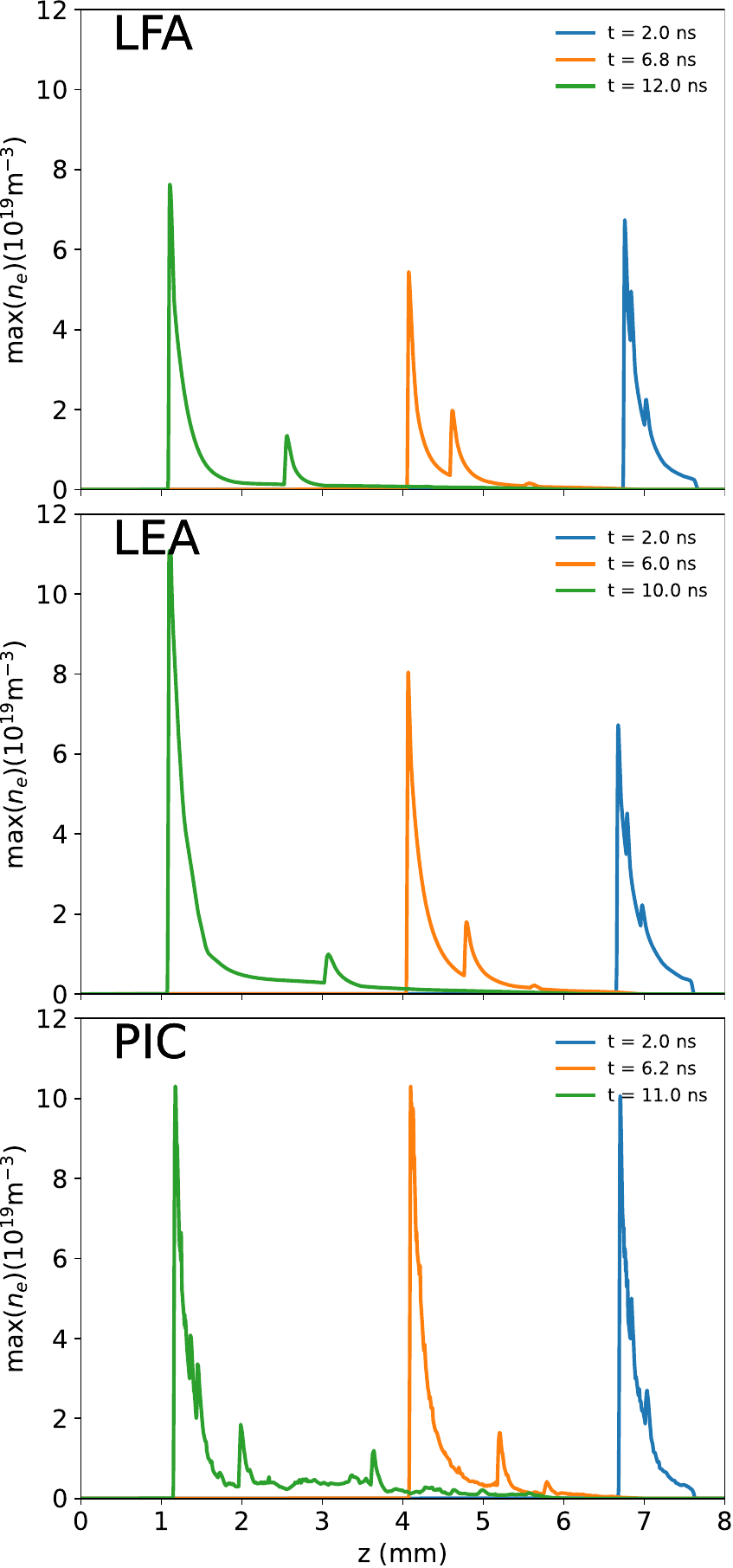}
        \caption{Maximum electron density}
        \label{fig:electron_density}
    \end{subfigure}\hfill
    \begin{subfigure}{0.49\linewidth}
        \centering
        \includegraphics[width=\linewidth]{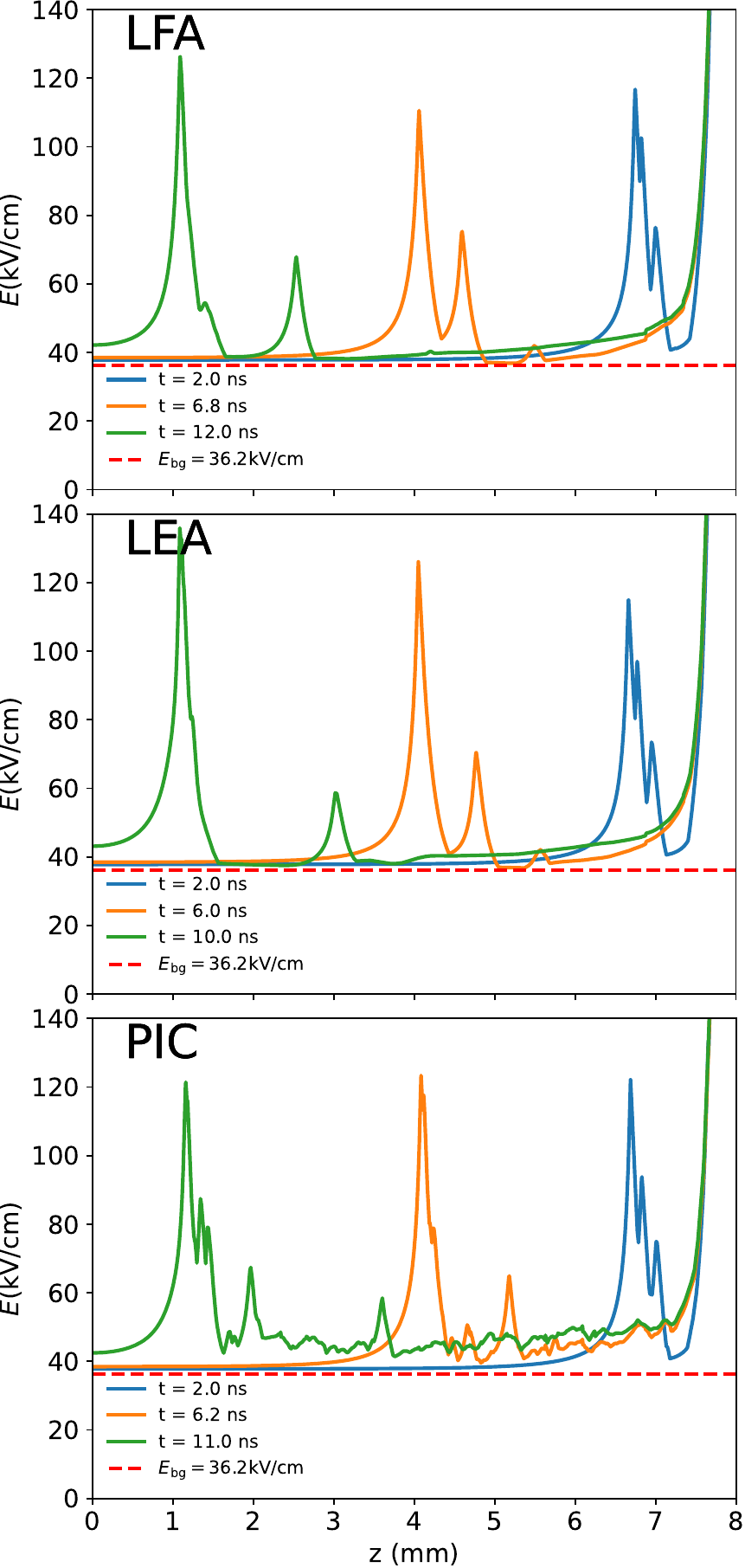}
        \caption{Maximum electric field}
        \label{fig:Electric_field}
    \end{subfigure}
    \caption{Comparison of \textbf{(a)} maximum electron density and \textbf{(b)} maximum electric field over the transverse direction as a function of the $z$-coordinate of the LFA, LEA, and PIC simulations in figure~\ref{fig:36kV}. 
    Three instances (marked by different colors) are chosen where the fastest streamers have roughly the same length. 
    The $z$-axis is truncated at $z = 8\;\mathrm{mm}$ to exclude the electrode. 
    The models produce comparable profiles, with PIC showing a noisier distribution.}
    \label{fig:subplots_density_field}
\end{figure*}

We first present simulations in CO$_2$ with an admixture of $1\%$ CFN in a background field of $36.1 \; \textrm{kV/cm}$ which is slightly above $E_k = 35.4$~kV/cm (see table \ref{tab:Critical_field}).
For these parameters we compare the simulation results of the LFA, LEA and particle models and show them in figures~\ref{fig:36kV} to~\ref{fig:subplots_density_field}.

Initially, the streamer morphologies and densities are highly similar, as shown in Figures \ref{fig:36kV} and \ref{fig:electron_density}.
After about $6\;\mathrm{ns}$, the main difference is that there is more branching in the particle model, as expected.
The particle model is stochastic in nature as particles scatter according to a Monte Carlo procedure, and stochastic density fluctuations are artificially enhanced due to the use of super-particles (see section~\ref{subsection:picmodel}).
Such dynamic density fluctuations are absent in the deterministic fluid models.

From figure \ref{fig:electron_density}, we have determined the velocities of the fastest streamer between $z \approx 7\,\mathrm{mm}$ and $z \approx 1 \, \mathrm{mm}$: $v_{\textrm{LEA}} \approx 0.70\;\mathrm{mm/ns}$, $v_{\textrm{LFA}} \approx 0.56\;\mathrm{mm/ns}$, and $v_{\textrm{PIC}} \approx 0.60\;\mathrm{mm/ns}$.
The LEA streamers are thus the fastest, while the LFA and PIC results agree rather well.
The higher velocity of the LEA streamers is related to their slightly higher maximal electric field and therefore higher electron drift velocity (as negative streamers should propagate at least with the drift velocity); see figure~\ref{fig:subplots_density_field}. 

The charge density $\rho$ at time $t = 8\;\mathrm{ns}$ is shown in figure \ref{fig:36kV_rho_all}.
The PIC simulations show a rather noisy charge density created by the many small streamer branches, whereas the charge density in the fluid simulations is much smoother.
The models start with a stochastic initial condition, which largely fades out in the fluid models.
The particle model remains stochastic, as it simulates the random motion of particles.
The stochastic effects will be discussed in more detail in section~\ref{subsection:add_stoch}

The electric field over the cross section is shown in figure~\ref{fig:36kV_electric_fld}.
In all models we observe `isolated' streamer heads, which only have a short conductive channel where the electric field is screened.
Further back, the field relaxes to the background field, which happens more smoothly in the fluid simulations.

The peak field is obtained by taking the maximum value of the electric field across the transverse direction as a function of the z-coordinate.
Figure \ref{fig:Electric_field} shows the peak field values.
In the PIC and LFA models, the peak electric field stays around $120\;\mathrm{kV/cm}$ in the whole gap, whereas this value increases to about $140\;\mathrm{kV/cm}$ in the LEA model when the streamer has almost bridged the gap.
In all cases, a main peak due to the fastest streamer is followed by smaller peaks corresponding to slower streamers.

\subsection{Comparison of model results in different fields}
\label{subsection:compare_diff_voltage}
\begin{figure}
    \centering
        \includegraphics[width=1\linewidth]{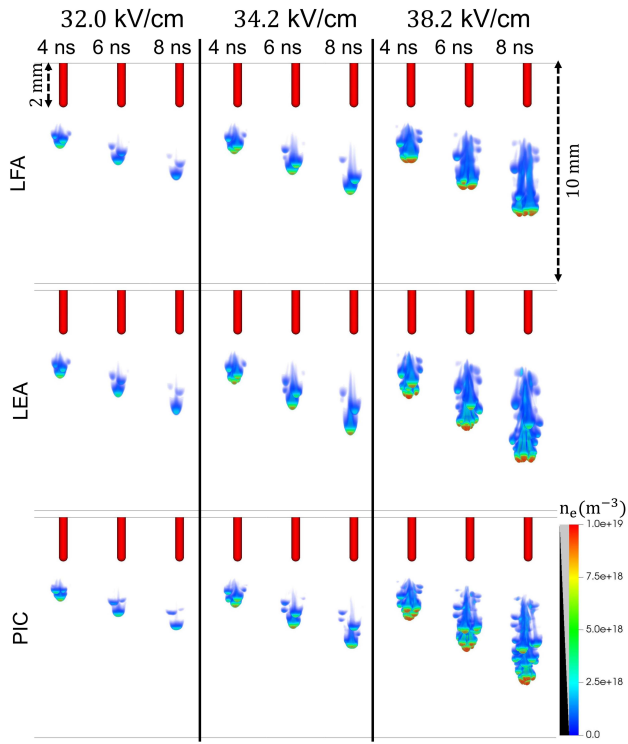}
        \caption{Evolution of the electron density with the same parameters as in figure~\ref{fig:36kV} except for the background field being here $E_{\mathrm{bg}}=32.0\;\mathrm{kV/cm}, \;34.2\;\mathrm{kV/cm}$ and $38.2\;\mathrm{kV/cm}$ with from left to right; in figure~\ref{fig:36kV} it was 36.1~kV/cm. Three time steps for simulations with the three models are shown. 
        At low $E_{\mathrm{bg}}$, the evolution is similar. At higher $E_{\mathrm{bg}}$, more branching occurs making the fluid and particle models less similar.}
        \label{fig:all_reduced}
\end{figure}

Figure~\ref{fig:all_reduced} shows simulation results in CO$_2$ with an admixture of 1\% CFN, but now at background fields of $E_{\mathrm{bg}} = 32.0,\;34.2$ and $38.2\;\mathrm{kV/cm}$, which are near $E_k = 35.4$~kV/cm.
With a higher background field the streamers have a longer conductive channel and there is significantly more streamer branching.
The particle simulations still branch more often, which is again the most prominent difference between fluid and particle simulations.
Similar to the case at $36.1$ kV/cm, streamers are the slowest with the LFA model and the fastest with the LEA model, with differences of 10\% to 20\%.

\subsection{Model comparison with 10\% CFN}
\label{subsection:different_concentration}
\begin{figure}
    \centering
    \includegraphics[width=1\linewidth]{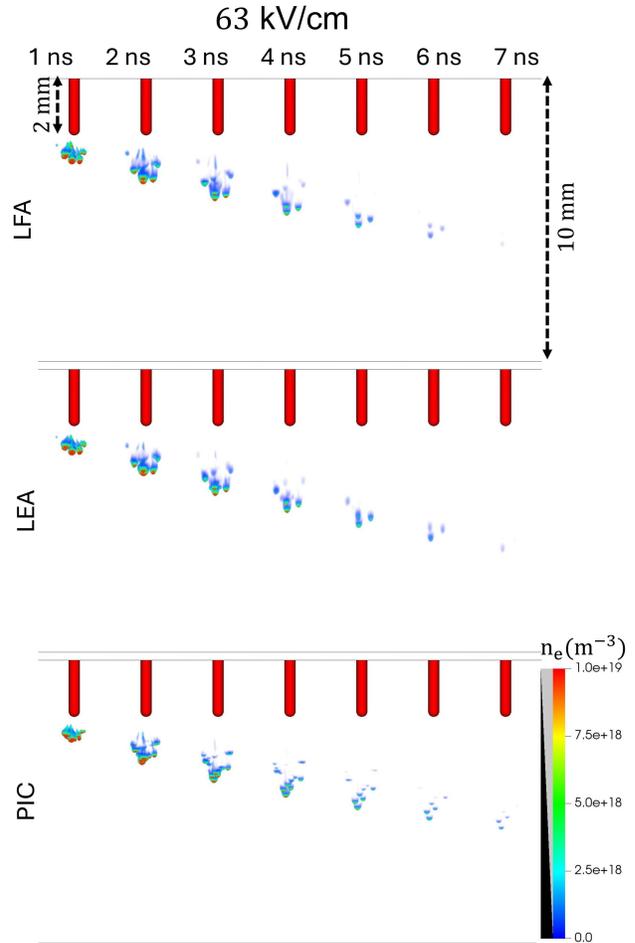}
    \caption{Evolution of the electron density, now in CO$_2$ with an admixture of $10\%$ XFN. The background field now is $E_{\rm bg}=63\;\mathrm{kV/cm}$, slightly lower than the critical field $E_k=69.3$~kV/cm. The models show similar initial growth across the different models, with some disagreement in the later phase of the simulation. All models predict streamers to fade.}
    \label{fig:63kV}
  \end{figure}

Next, we consider an admixture of 10\% CFN at $63\;\mathrm{kV/cm}$ which is below $E_k$, see table \ref{tab:Critical_field}.
The results of the LFA, LEA and particle models are shown in figure ~\ref{fig:63kV}.
With all models, we observe streamer heads with a short conductive channel that fade out after propagating approximately halfway through the gap.
The LEA streamers are again the fastest, with the LFA and PIC velocities agreeing well, similar to the cases with a lower CFN fraction.

\subsection{Larger computational domain}
\label{subsection:big_domain}
 \begin{figure*}
    \centering
    \includegraphics[width=0.7\textwidth]{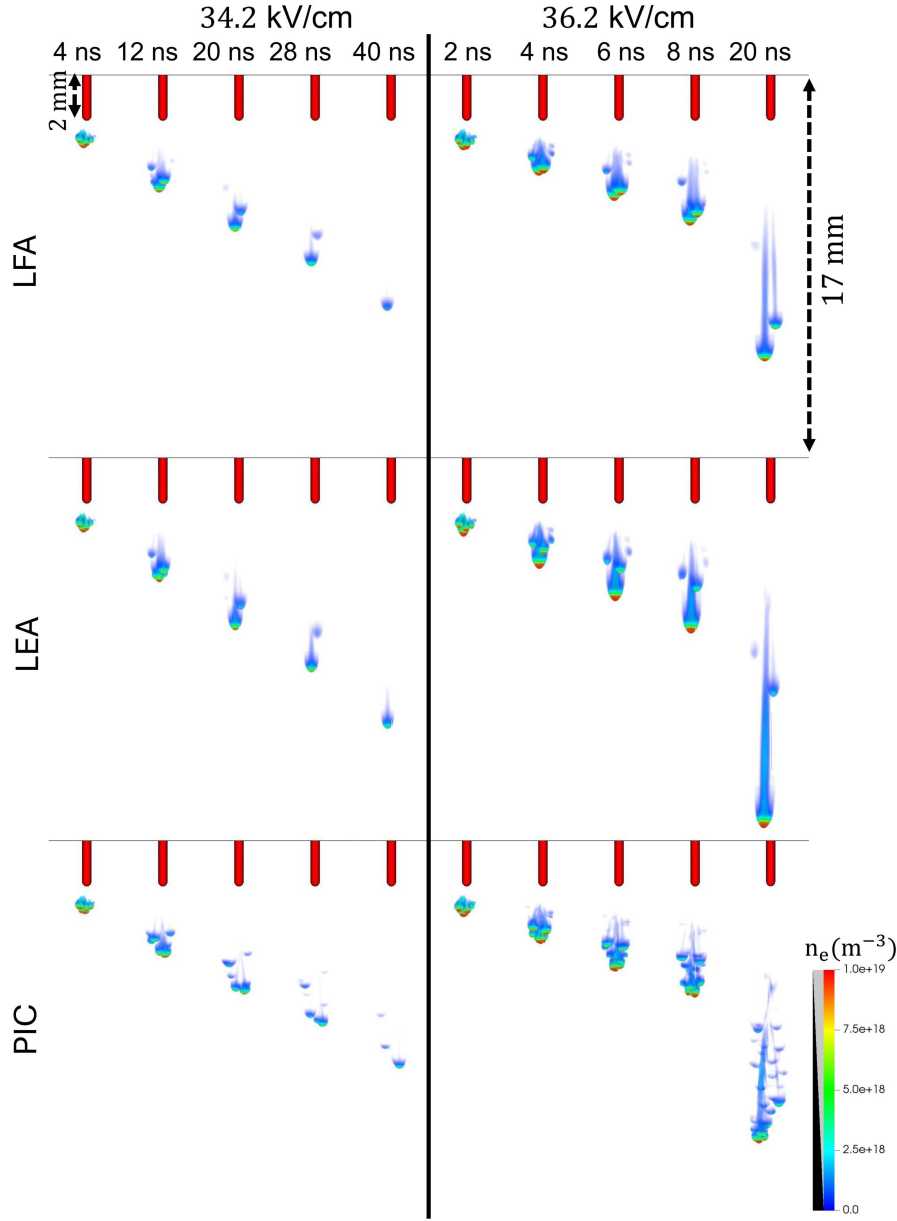}
    \caption{Evolution of the electron density within the larger domain in CO$_2$ with an admixture of $1\%$ CFN in a background field of $E_{\mathrm{bg}} = 34.2\;\mathrm{kV/cm}$ (left) and $36.1\;\mathrm{kV/cm}$ (right). Again, similar streamers are produced with the different models.}
    \label{fig:big_domain}
\end{figure*}

The simulations presented above were performed in a relatively small computational domain of $5\;\mathrm{mm}\times 5\;\mathrm{mm}\times 10\;\mathrm{mm}$.
We have also simulated streamers in a larger domain of $10\;\mathrm{mm}\times10\;\mathrm{mm}\times20\;\mathrm{mm}$ while keeping the same needle size to reduce the effect of the lateral boundary conditions.
The simulations were performed with an admixture of $1\%\;\mathrm{CFN}$ at $E_{\mathrm{bg}} = 34.2\;\mathrm{kV/cm}$ and $36.1\;\mathrm{kV/cm}$.
Results of the LFA, LEA and particle models are shown in figure ~\ref{fig:big_domain}.

At $34.2\;\mathrm{kV/cm}$ all models produce isolated streamer heads with only a short channel.
These streamers stop propagating at about 5 mm below the electrode.
Within the smaller domain the streamer crossed the gap at this background field, see figure~\ref{fig:all_reduced}.
However, it should be noted that we define the background field as the average field between the plates.
When the background fields are equal, the \emph{average} field between the electrode tip and the opposite plate will be lower in the larger computational domain.

At $E_{\mathrm{bg}} = 36.1\;\mathrm{kV/cm}$ the streamers have much longer conductive channels.
This causes streamers to accelerate while they cross the gap, particularly for the LEA model, which again produces the fastest streamers, whereas the PIC and LFA velocities agree quite well.
As before, there is more branching in the PIC simulations.

%%% Local Variables:
%%% mode: LaTeX
%%% TeX-master: "main"
%%% End:

\section{Discussion}
\label{section:chapter4}
\subsection{Computational cost}
\label{subsection:cost}
The simulations were performed on ``Rome'' computing nodes of the Snellius supercomputer (SURF, the Netherlands), which have dual AMD EPYC 7H12 CPUs with a total of 128 cores.
For each simulation a quarter (32 cores) of a single node was used.
The approximate run times of the simulations in~\ref{subsection:different_bg_fld} and~\ref{subsection:compare_diff_voltage} are summarised in table~\ref{table:run_time}.

The particle model only refined grid cells in which $n_e \geq 10^{13}\;\mathrm{m}^{-3}$, so it generally used fewer grid cells than the fluid models.
At low $E_{\mathrm{bg}}$, the use of relatively few super-particles (around $10^7$) resulted in PIC having the shortest run times.
At higher $E_{\mathrm{bg}}$, the number of super-particles increased significantly (up to $10^8$), which slowed down PIC, and the LFA became faster. 
The LEA model was always the slowest, which was caused by two main factors: a finer grid due to the slightly higher electric field at streamer tips, and a smaller time step due to the extra energy equation.

\begin{table}
\centering
\caption{The approximate simulation times of the runs in section~\ref{subsection:different_bg_fld} and~\ref{subsection:compare_diff_voltage}.}
\label{table:run_time}
\begin{tabular}{c| c c c}
    \hline
    $\mathbf{E_{bg}}$ & \textbf{LFA} & \textbf{LEA} & \textbf{PIC} \\
    \hline
    32.0 kV/cm   & $\sim19\;\mathrm{h}$ & $\sim65\;\mathrm{h}$  & $\sim5\;\mathrm{h}$ \\
    34.2 kV/cm & $\sim36\;\mathrm{h}$ & $\sim64\;\mathrm{h}$  & $\sim18\;\mathrm{h}$ \\
    36.1 kV/cm & $\sim25\;\mathrm{h}$ & $\sim84\;\mathrm{h}$  & $\sim44\;\mathrm{h}$ \\
    38.2 kV/cm & $\sim34\;\mathrm{h}$ & $\sim108\;\mathrm{h}$ & $\sim79\;\mathrm{h}$ \\
    \hline
\end{tabular}
\end{table}

\subsection{Sensitivity to initial conditions}
\label{subsection:ic_sensitivity}
\begin{figure}
  \centering
  \includegraphics[width=\linewidth]{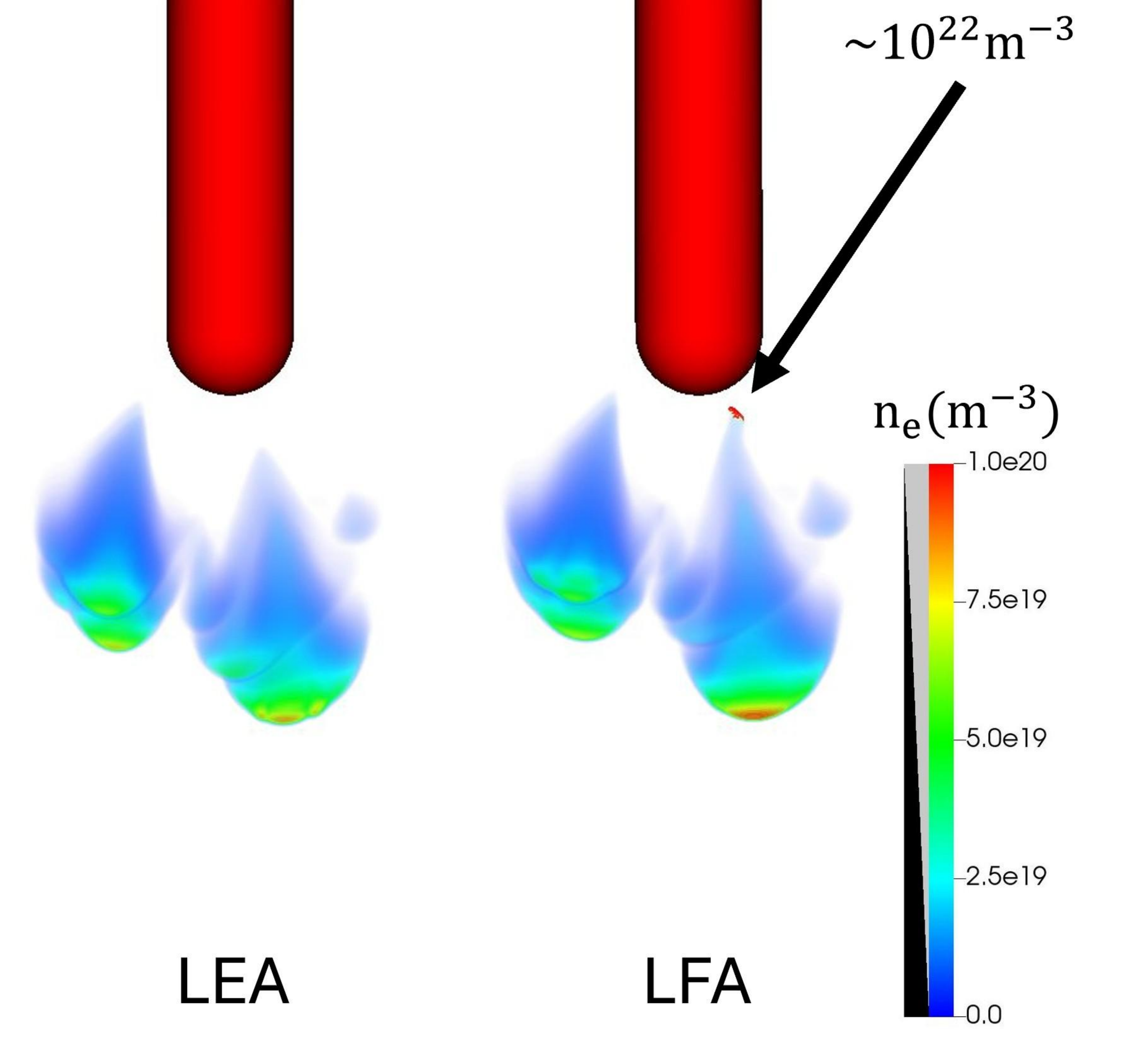}
  \caption{A close-up of the electron density near the electrode with the seed placed closer to the electrode. In LFA a tail moving backwards towards the electrode appears, which does not appear in LEA for the considered parameters. The density in this tail quickly increases due to fast ionisation, effectively stopping the LFA simulation, as the required time step keeps decreasing.}
  \label{fig:ic_issue}
\end{figure}
As is shown in section~\ref{subsection:fluidmodel} and in~\cite{marskar20243d}, the LFA model can have issues resolving the cathode sheath.
The sheath issue is illustrated in figure \ref{fig:ic_issue}, which shows LEA and LFA simulations in which the initial electrons were shifted $0.1\;\textrm{mm}$ towards the electrode tip in comparison to the simulations shown earlier in section~\ref{subsection:different_bg_fld}.

With the LFA model, the back side of a negative streamer grows towards the needle, resulting in a small region with an electron density as high as about $10^{22}\;\textrm{m}^{-3}$.
This high density keeps increasing, and due to the resulting small dielectric relaxation time $\Delta t < \epsilon_0/(e\mu_en_e)$ \cite{Teunissen_2017} the simulation effectively stops.
Such unphysical growth is not present in the LEA model under the conditions and time scale considered here.

Note that in our LFA simulations, we have used a correction factor for the source term, as shown in equation~\eqref{eq:source_correction}.
This factor reduces the unphysical growth, but does not completely stop it, as is evident in figure~\ref{fig:ic_issue}.
In general, one should therefore avoid initial conditions that lead to a cathode sheath when using the LFA model.

\subsection{Neumann versus Dirichlet boundary conditions}
\label{subsection:BC_versus}

Commonly a Neumann zero boundary condition ($\mathbf{n}\cdot\nabla n_e=0$) is used for $n_e$ on the electrode for positive streamers. 
In this case, electrons flow into the electrode from which the streamers are emitted. 
However, for negative streamers electrons can flow out of the electrode, making the boundary conditions more important.
We therefore study the effect of Dirichlet or Neumann boundary conditions (B.C.s) for the electron density at the electrode.
We use CO$_2$ with an admixture of $1\%$ CFN in a background field of $36.1\;\mathrm{kV/cm}$ in the LFA model.

\begin{figure}
    \centering
    \includegraphics[width=\linewidth]{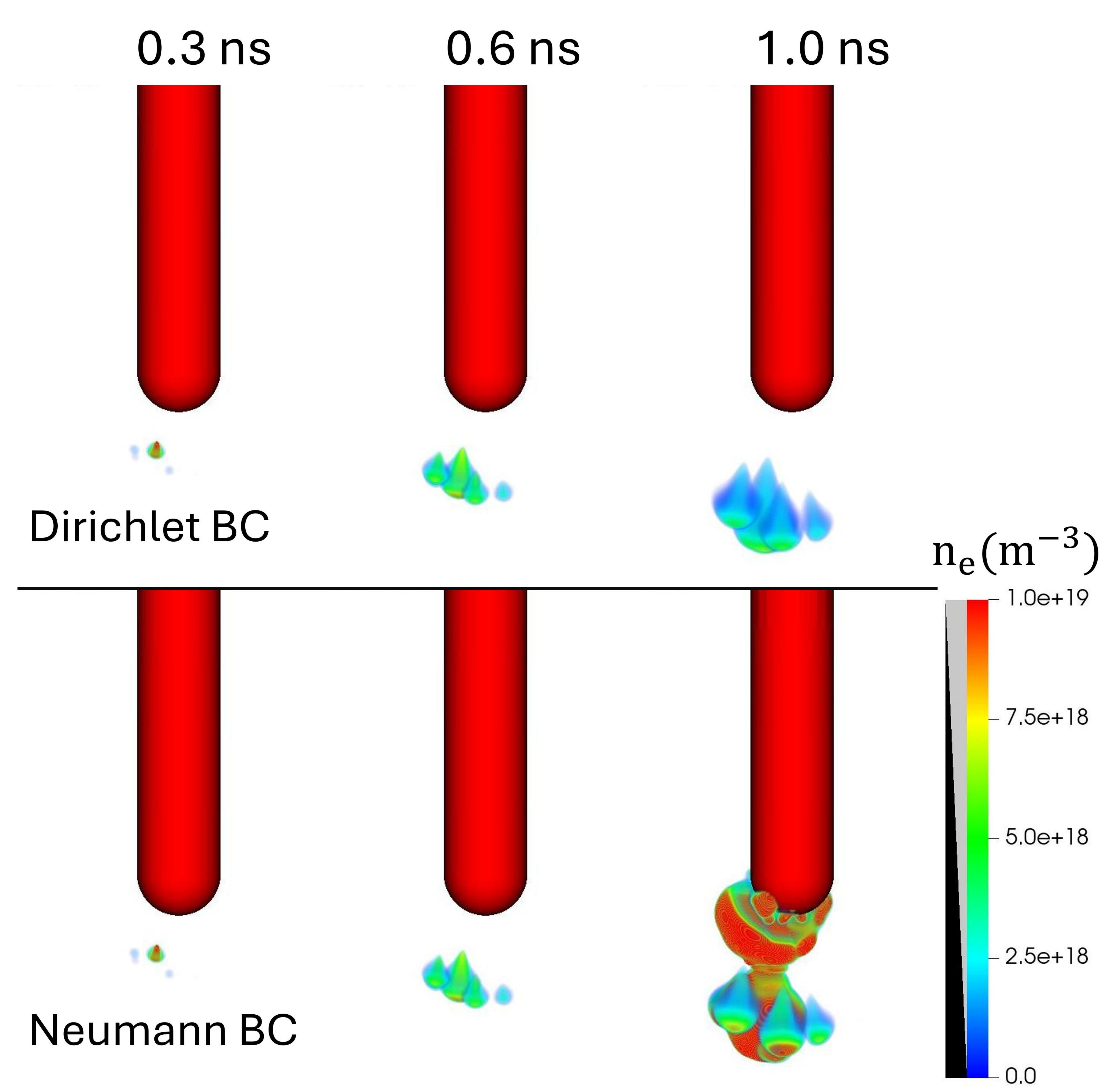}
    \caption{The evolution of the electron density near an electrode with either Dirichlet (top) or Neumann (bottom) zero boundary conditions using the same parameters as in~\ref{subsection:compare_diff_voltage} s.t. $E_{\mathrm{bg}}=36.1\;\mathrm{kV/cm}$. Neumann conditions allow the electrons to flow out of the electrode and to form structures of high density around the electrode effectively stopping the simulation. 
    }
    \label{fig:36kV_BC}
\end{figure}

A close-up of the electron dynamics near the electrode is shown in figure~\ref{fig:36kV_BC}.
In the first two stages of figure \ref{fig:36kV_BC} the number of initial avalanches and their morphology and electron densities are rather indistinguishable between the two B.C.s.
The difference emerges after $0.6\;\mathrm{ns}$ as in the case of the Neumann B.C., structures with a high electron density form near the electrode.
These high densities result in a short dielectric relaxation time, which limits the global time step.
These high electron density structures form after a small electron density has diffused towards the electrode. 
Most of these electrons drift away, causing the flux at the electrode to be non-zero. 
This results in an outflow of electrons with Neumann zero B.C..
The combination of this outflow with the local high electric field causes ionisation, resulting in rapidly growing $n_e$ effectively stopping the simulation.
This outflow is absent with a Dirichlet B.C., preventing high $n_e$ formation in the initial phase. 
Dirichlet zero B.C. approximates a scenario where electrons do not have sufficient energy to overcome the electrode work function, such that there is no electron emission. 
The short timescales of 20 ns considered in this work result in a small ion flux and low secondary electron emission, thus Dirichlet B.C. are more suitable.

\subsection{Stochastic effects}
\label{subsection:add_stoch}

The initial condition for the electrons is stochastic, as described in section~\ref{subsubsection:fluid_initial_and_boundary_conditions}. 
The dynamic evolution of the fluid models is deterministic, while the particle model is stochastic. 
Thus, it is understandable that the particle model branches more frequently than fluid models, as shown in section~\ref{section:chapter3_2}.
Branching can occur due to physical density fluctuations.
The particle model overestimates density fluctuations due to the use of super-particles, where the fluctuations are increased by a factor of $\sqrt{w}$, with $w$ the particle weights \cite{nijdam2020physics}.
For the $36\;\mathrm{kV/cm}$ case in section~\ref{subsection:different_bg_fld} the average weight of the simulations is around 25, such that fluctuations are artificially increased by about a factor five.
On the contrary, fluid models underestimate fluctuations.
Thus, PIC sets an upper bound and fluid models a lower bound to the discharge stochasticity.
PIC with particle weight 1 would give correct fluctuations, but is not feasible due to the large number of particles required. 

\subsection{Streamer polarity and sources of free electrons}
\label{subsection:free_electron_sources}

This work focused on negative streamers, which can propagate without nonlocal sources of free electrons, such as photoionisation or electron detachment. 
Photoionisation is an important mechanism for streamers in air, allowing positive streamers to propagate against the electron drift direction.
The potential sources of free electrons in CO$_2$ with admixtures of CFN remain uncertain.

In \cite{li2024investigation, marskar20243d}, photoionisation was implemented in pure $\textrm{CO}_2$.
It was found that the weak photoionisation of CO$_2$ could sustain the propagation of positive streamers in sufficiently high background fields.
There was however considerable uncertainty in the photoionisation parameters.
Photoionisation in mixtures of CFN and CO$_2$ could be approximated by the photoionisation parameters for pure CO$_2$, but energetic photons might efficiently be absorbed by CFN.
Such an effect was found in~\cite{bouwman20223d}, where it was shown that adding CH$_4$ to air leads to a reduction in photoionisation.

Another source of free electrons could be electron detachment in a pre-ionized gas.
Previous experimental work on swarm experiments shows evidence of detachment in $\textrm{CFN}$ \cite{hosl2019identification,rankovic2020temporary,mirpour2022investigating}.
The detachment reactions are given in \cite{rankovic2020temporary}, and the detachment reaction rates in \cite{hosl2019identification}.
However, the presence and density of negative ions in front of the discharge remain uncertain and will depend in particular on the experimental conditions and the previous discharge history.

%%% Local Variables:
%%% mode: LaTeX
%%% TeX-master: "main"
%%% End:

\section{Conclusion and Outlook}
\label{section:chapter5}
\subsection{Conclusion}
We first reviewed the transport and reaction coefficients in CO$_2$ with admixtures of CFN. 
The sensitivity of these coefficients to the choice of solver was analysed using BOLSIG+ and \texttt{particle\_swarm} for different CFN-CO$_2$ mixtures.
Both solvers produced similar $\mu N$, $\alpha/N$, and $\eta/N$, while larger deviations were found for $DN$ due to shortcomings of the two-term approximation \cite{hagelaar2025beyond}.
Overall, both solvers are suitable.

Next, we compared \texttt{particle\_swarm} results for transport and reaction coefficients with experimental data using different cross section sets at $1\%$ and $10\%$ CFN. 
Good agreement was obtained for $\mu N$ and $\alpha_{\mathrm{eff}}/N$ when using the Hayashi database for CO$_2$ combined with either the Flynn or Zhang datasets for CFN. 
Since these coefficients are the important inputs for fluid models, we chose the Hayashi set for CO$_2$ in our simulations.
Different CFN cross sections had only a minor influence on the mixture coefficients, and we could not determine whether the Flynn or the Zhang set showed better agreement.

Finally, we investigated whether fluid models can model stochastic negative CFN-CO$_2$ streamers in various background fields.
The fluid models approximate the particle model reasonably well, though branching occurred more frequently in particle simulations due to their intrinsic stochasticity.
The particle model overestimates fluctuations due to the use of super-particles, whereas fluid models underestimate them by averaging out fluctuations. 
Furthermore, with the LEA model streamers were generally the fastest.

\subsection{Outlook}
Combining modelling and experiments is essential for understanding the discharge behaviour and insulation performance of CFN-CO$_2$ mixtures.
An essential next step is to validate the fluid and particle models for negative CFN-CO$_2$ streamers through experiments, as has been done for positive air streamers~\cite{li2021comparing}.
Such a validation enables the use of models to study the current open questions, such as:
\begin{enumerate}
    \item What are the generic properties of streamers that have strong attachment and little to no photoionisation?
    \item Positive air streamers propagate more easily than negative streamers.
    Is this a generic streamer property, or will this be different for eco-friendly gases?
\end{enumerate}

Another important step is to extend the fluid and particle models with processes such as photoionisation, detachment, and three-body reactions. 
However, reliable data must be obtained first.
The CFN-CO$_2$ streamers are suspected to be highly stochastic, consisting of many smaller streamers.
Therefore, computationally expensive 3D simulations are necessary.
Developing reduced models as in e.g.~\cite{teunissen2025data,luque2014growing} might therefore be required.

%%% Local Variables:
%%% mode: LaTeX
%%% TeX-master: "main"
%%% End:

\section*{Acknowledgments}
The research of T.S.\ was supported by the Dutch Research Council (NWO) through the AES project 20344 ‘Green Sparks'. 
The present study used the Dutch national e-infrastructure with the support of the SURF Cooperative using grant no. EINF-10838.

%\newpage

\appendix
\label{section:appendix}
\section{Comparing CO$_2$ coefficients}
\label{appendix:co2_coefficients}
\begin{figure}
    \centering
    \includegraphics[width=0.9\linewidth]{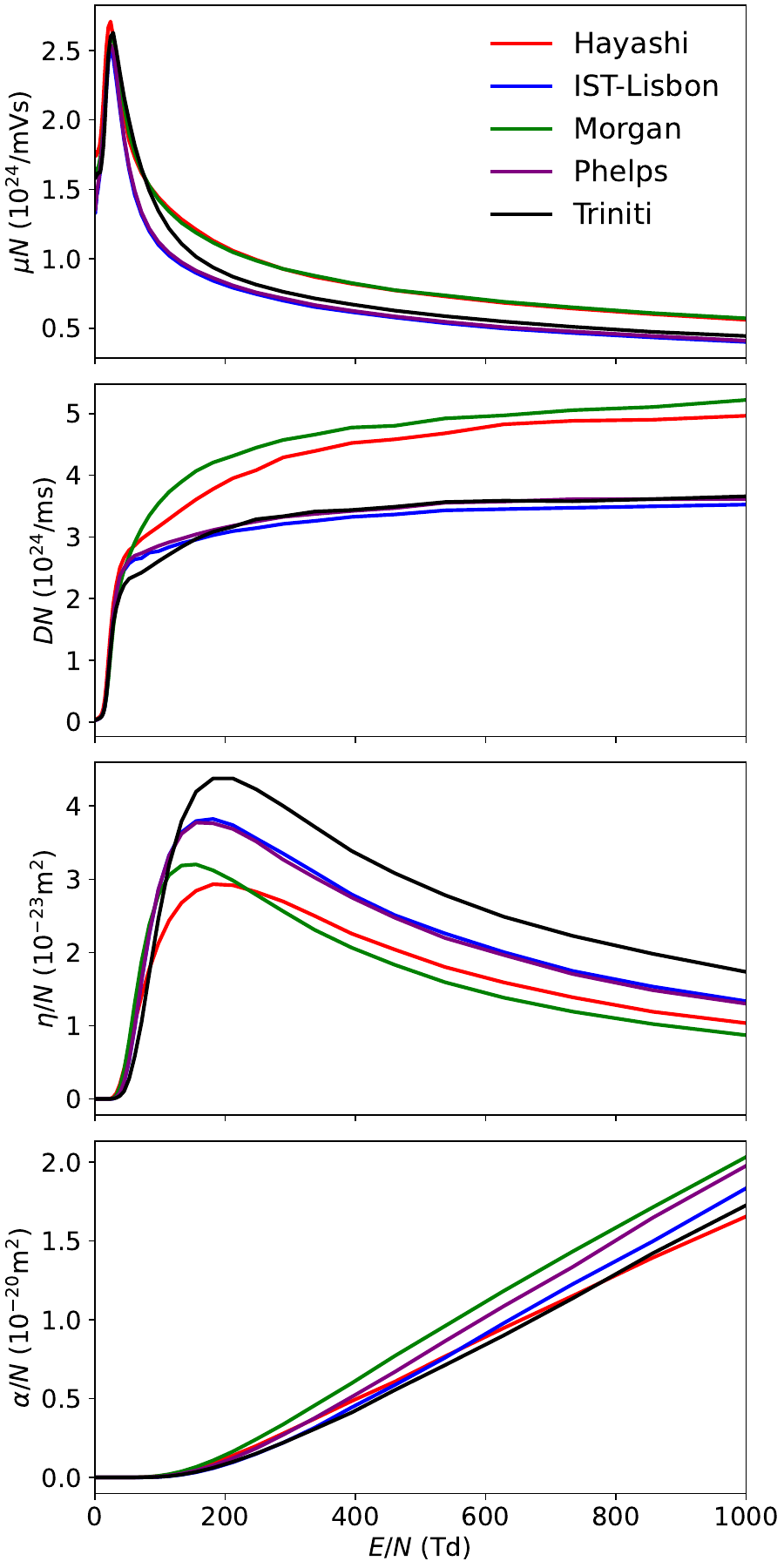}
    \caption{Comparison of the Hayashi, IST-Lisbon, Morgan, Phelps and Triniti reduced transport and reaction coefficients in pure CO$_2$ as function $E/N$.
    The coefficients are calculated with the particle\_swarm code.}
    \label{fig:co2_transport}
\end{figure}

The reduced transport and reaction coefficients are shown in figure~\ref{fig:co2_transport} of the Hayashi~\cite{Hayashi_co2}, IST-Lisbon~\cite{IST_co2}, Morgan~\cite{Morgan_co2}, Phelps~\cite{Phelps_co2} and Triniti~\cite{Triniti_co2} CO$_2$ cross sections. 
The particle$\_$swarm solver is used.
The calculated coefficients have a clear division between the Hayashi and Morgan databases and the other databases for $\mu N, DN$ and $\eta/N$.
We see this division less for $\alpha/N$.
Additionally, the values of the attachment rate $\eta/N$ vary between the IST-Lisbon, Phelps, and Triniti databases, whereas the Hayashi and Morgan databases give quite similar $\eta/N$ profiles.

%%% Local Variables:
%%% mode: LaTeX
%%% TeX-master: "main"
%%% End:

%\newpage

\section*{References}
\bibliographystyle{iopart-num}
\bibliography{References}

\end{document}